\documentclass[letterpaper, 10 pt, conference]{ieeeconf}
\IEEEoverridecommandlockouts 
\usepackage{amsmath,amsfonts,amssymb}
\usepackage{algorithmic}
\usepackage{algorithm}
\usepackage{array}
\usepackage{textcomp}
\usepackage{stfloats}
\usepackage{url}
\usepackage{verbatim}
\usepackage{multirow}
\usepackage{graphicx}
\usepackage{cite}
\usepackage[capitalise]{cleveref}
\usepackage{epsfig}
\usepackage{textcomp}
\usepackage{xcolor}
\usepackage{cases}
\usepackage[justification=centering]{caption}
\usepackage{subcaption}

\DeclareMathOperator*{\argmin}{arg\,min} 

\graphicspath{{Images/}}
\title{\LARGE \bf
Constrained Control of PDE Traffic Flow via Spatial Control Barrier Functions
}
\author{Brian Block and Stephanie Stockar
\thanks{This material is based upon work supported by the National Science Foundation CAREER Award 2042354.}
\thanks{The authors are with the Department of Mechanical and Aerospace Engineering and the Center for Automotive Research, The Ohio State University, 930 Kinnear Road, Columbus, OH 43212 USA.}}

\begin{document}

\maketitle
\thispagestyle{empty}
\pagestyle{empty}

\begin{abstract}
In this paper, a constrained control approach to variable speed limit (VSL) control for macroscopic partial differential equation (PDE) traffic models is developed. Control Lyapunov function (CLF) theory for ordinary differential equations (ODEs) is extended to account for spatially and temporally varying states and control inputs. The stabilizing CLF is then unified with safety constraints through the introduction of spatially varying control barrier functions (sCBFs). These methods are applied to in-domain VSL control of the Lighthill–Whitham–Richards (LWR) model to regulate traffic density to a desired profile while ensuring the density remains below prescribed limits enforced by the sCBF. Results show that incorporating constrained control minimally affects the stabilizing control input while successfully maintaining the density within the defined safe set.
\end{abstract}

\section{INTRODUCTION}
In traffic systems management, reducing congestion and regulating the system toward a desired traffic density profile has been the focus of numerous control strategies in literature \cite{block_lq_2024,yu_traffic_2019,yu_reinforcement_2022}. However, it is equally critical to ensure that the traffic state remains within safe or acceptable bounds \cite{siri_freeway_2021,pasquale_comparative_2016}. This includes preventing density from exceeding a prescribed maximum and avoiding excessively high or low velocities. Accordingly, constraint handling must be incorporated directly into traffic control strategies.

Traffic is generally modeled via partial differential equations (PDE) that describe the density and flow of vehicles through the system \cite{siri_freeway_2021}. For in-domain control of PDE-based traffic models, variable speed limit (VSL) control is frequently used to modify the speed limit and, therefore, influence vehicle flow \cite{carlson_local_2011,muralidharan_computationally_2015,karafyllis_feedback_2019-1,block_lq_2024,block_lq-informed_2025}. In many approaches, the PDE is discretized into a system of ordinary differential equations (ODEs), and the controller is designed on the discrete model. For example, \cite{carlson_local_2011} develops a discrete feedback controller to mitigate bottlenecks, while \cite{muralidharan_computationally_2015} formulates a model predictive controller to reduce congestion. Alternatively, the controller can be designed directly in the PDE domain \cite{karafyllis_feedback_2019-1,block_lq_2024,block_lq-informed_2025}. In \cite{karafyllis_feedback_2019-1}, a feedback law is constructed to regulate the LWR model to a desired density profile, and \cite{block_lq_2024} develops an optimal controller based on a linear quadratic regulator to achieve the same objective. This approach is further used in \cite{block_lq-informed_2025} to generate an implementable discretized rule-based controller. However, these methods primarily focus on regulation of the density to a desired profile and do not explicitly enforce state constraints or guarantee that the system remains within a prescribed operating range.

The requirement to keep a system within a designated set is typically formalized through the forward invariance of that set \cite{ames_control_2014,ames_control_2017,ames_control_2019}, which is achieved through the use of control barrier functions (CBFs). The use of CBFs for safety-critical control emerged from their unification with stabilizing control Lyapunov function (CLF) theory \cite{wieland_constructive_2007,ames_control_2019}. For systems described by ODEs, particularly in automotive control applications, CBFs have been applied to ensure safe operation of adaptive cruise controllers \cite{ames_control_2014,ames_control_2017,xiao_control_2019}. In \cite{ames_control_2014}, CLFs and CBFs are unified under a quadratic programming (QP) optimization problem that enforces both desired vehicle speed and force at the wheel constraints. This methodology is extended in \cite{ames_control_2017} to incorporate zeroing and reciprocal barrier functions, and subsequently applied to adaptive cruise control and lane-keeping. Higher-order CBFs are used in \cite{xiao_control_2019} to address similar cruise control problems. An event-triggered control strategy is proposed in \cite{sabouni_optimal_2024} to address feasibility issues in safe connected and automated vehicle (CAV) control.

While the application of barrier functions to ODE systems is well established, their use in PDE systems remains underexplored. Barrier functionals have been used, for example, in \cite{ahmadi_barrier_2015,ahmadi_safety_2017} to estimate outputs of PDE systems. However, the application focused on verification of constraint satisfaction rather than control design. In \cite{park_discretization-robust_2023}, the Euler–Bernoulli beam equation is discretized and a CBF-based controller is applied to the resulting ODE approximation rather than directly to the PDE. A boundary feedback controller is developed in \cite{karafyllis_global_2022} to guarantee a lower bound on the density of a compressible flow, though the combined use of CLFs and CBFs was not investigated.  In \cite{koga_event-triggered_2023,koga_safe_2023}, the boundary control of a one-dimensional Stefan PDE for the melting of a material was obtained using CBFs and event-triggered control. This approach, though, used a finite-dimensional approximation of the continuous control input.

This paper presents a unified control framework that introduces spatially and temporally varying CBFs for in-domain control of PDEs and provides a novel formulation enabling safety-critical constraint enforcement directly at the PDE level. Furthermore, a CLF-based controller is unified with the proposed CBF-based controller within a quadratic programming framework to simultaneously achieve regulation to a desired profile and enforcement of state constraints in infinite-dimensional systems.

The paper is structured as follows. First, in \cref{sec:ModelDescription} the traffic PDE is described and the control design for VSLs is introduced. Then in \cref{sec:CLF}, CLF theory is extended for use in PDE systems and in \cref{sec:CBF} CBF theory is extended to a spatially varying control input and combined into a quadratic programming problem. Finally, in \cref{sec:Results} the developed controllers are validated on a case study.

\section{MODEL DESCRIPTION}\label{sec:ModelDescription}
\subsection{Lighthill-Whitham-Richards Model}
\begin{figure}[thpb]
  \centering
  \includegraphics[width=1\columnwidth]{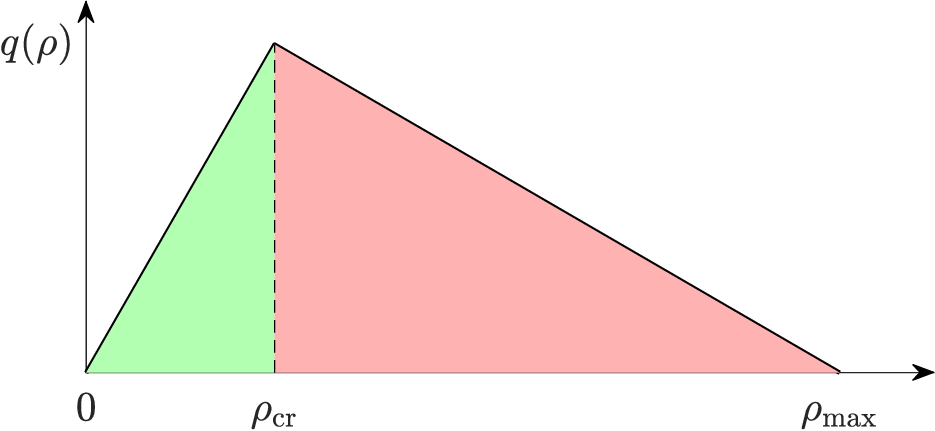}
  \caption{Triangular fundamental diagram.}
  \label{fig:fd}
\end{figure}

The LWR model \cite{lighthill_kinematic_1955, richards_shock_1956} is a first-order, hyperbolic conservation equation given by
\begin{equation} \label{eq:LWR}
    \frac{\partial\rho(z,t)}{\partial t} + \frac{\partial q(\rho(z,t))}{\partial z} = 0
\end{equation}
where $\rho(z,t)$ is the traffic density, representing the number of vehicles occupying a road length, $q(z,t)$ is the traffic flow rate, and $z\in\mathbb{R}^n$ and $t\in\mathbb{R}^m$ are the space and time variables, respectively. The closure of the problem is achieved through the adoption of a fundamental flow diagram that relates density with flow. The fundamental diagram used in this paper is the triangular model \cite{daganzo_cell_1994}, shown in \cref{fig:fd} and described by
\begin{equation} \label{eq:fd}
    q(\rho(z,t)) = 
    \begin{cases} V_\mathrm{max}\rho(z,t), & \forall \rho(z,t)\leq\rho_\mathrm{cr}\\
                  w(\rho_\mathrm{max} + \rho(z,t)), & \forall \rho(z,t)>\rho_\mathrm{cr}        
    \end{cases}
\end{equation}
where $V_\mathrm{max}$ is the maximum velocity, $w$ is the congested speed, $\rho_\mathrm{cr}$ is the critical density, and $\rho_\mathrm{max}$ is the maximum density of the road. The congested speed $w$ is given as
\begin{equation} \label{eq:w}
    w = \frac{V_\mathrm{max}\rho_\mathrm{cr}}{\rho_\mathrm{max}-\rho_\mathrm{cr}}
\end{equation}

\subsection{Variable Speed Limits}
The variable speed limit actuation is included in the governing equations by modifying the fundamental diagram in \cref{eq:fd} as
\begin{equation} \label{eq:fdVSL}
    q(\rho(z,t),b(z,t)) = 
    \begin{cases} b(z,t)V_\mathrm{max}\rho(z,t), & \forall \rho\leq\rho_\mathrm{cr}\\
                  b(z,t)w(\rho_\mathrm{max} + \rho(z,t)), & \forall \rho>\rho_\mathrm{cr}       
    \end{cases}
\end{equation}
where $b(z,t)$ is the VSL ratio. A value of $b(z,t)>1$ increases the slope of both the free flow and congested regime in \cref{fig:fd}, while a value of $b(z,t)<1$ will decrease it. This, in turn affects the maximum possible flow $q_\mathrm{max}$, which occurs at the critical density. For simplicity, for the rest of the paper $(z,t)$ is omitted, except where it is needed for emphasis.

\subsection{Linearization of LWR Model}
The linearized LWR model is derived considering the nominal equilibrium density $\rho_0$ and nominal maximum velocity ratio $b_0$ \cite{block_lq_2024} such that the variables can be written as
\begin{equation}\label{eq:linearizedVSLfd}
    \begin{aligned}
        \rho &= \rho_0 + \Delta\rho\\
        b &= b_0 + \Delta b\\
        q &=
        \begin{cases}
            v_\mathrm{max}((\rho_0 + \Delta\rho)(b_0 + \Delta b)), & \rho\leq\rho_\mathrm{cr}\\
            w((\rho_\mathrm{max} - \rho_0 - \Delta\rho)(b_0 + \Delta b)), & \rho>\rho_\mathrm{cr}
        \end{cases}
    \end{aligned}
\end{equation}
The LWR model \cref{eq:LWR} can then be written in standard, linear hyperbolic PDE form
\begin{equation} \label{eq:standardlinearPDE}
    \begin{aligned}
        \frac{\partial x}{\partial t}(z,t) &= \mathcal{V}\frac{\partial x}{\partial z}(z,t) + \mathcal{M}x(z,t) + Bu(z,t) \\
        y(z,t) &= Cx(z,t)
    \end{aligned}
\end{equation}
where
\begin{equation} \label{eq:LWRLQRvariables}
    \begin{aligned}
        \mathcal{V} &= \begin{cases} -V_\mathrm{max}b_0, & \forall \rho\leq\rho_\mathrm{cr}\\
                           wb_0, & \forall \rho>\rho_\mathrm{cr}        
             \end{cases}\\
        \mathcal{M} &= 0\\
        B &= \begin{cases} -V_\mathrm{max}\rho_0, & \forall \rho\leq\rho_\mathrm{cr}\\
                           -w(\rho_\mathrm{max}-\rho_0), & \forall \rho>\rho_\mathrm{cr}        
             \end{cases}\\
        C &= I\\
        x &= \Delta\rho,\quad u = \frac{\partial\Delta b}{\partial z}
    \end{aligned}
\end{equation}
The control input $u=\frac{\partial\Delta b}{\partial z}$ is the change in the variable speed limit over the length of road.

\section{STABILIZING CONTROL OF TRAFFIC MODELS}\label{sec:CLF}

\subsection{Control Lyapunov Theory for ODEs}
To motivate the description of safety for a system, first it is necessary to discuss the family of stabilizing controllers. Consider the system
\begin{equation}\label{eq:controlaffinesystem}
    \frac{dx}{dt} = f(x) + g(x)u
\end{equation}
where $f(x)$ and $g(x)$ can be either nonlinear or linear functions of the state $x$. If the goal is to stabilize the system \cref{eq:controlaffinesystem} to a point $x^*=0$ or equivalently $x(t) \rightarrow 0$, this is equivalent to finding a feedback control law that drives a positive definite function, $V : D\subset \mathbb{R}^m \rightarrow \mathbb{R}_{\geq 0}$, to zero \cite{ames_control_2019}. That is, if
\begin{equation}\label{eq:defstabcontrol}
    \exists\, u=k(x) \quad \mathrm{s.t.} \quad \Dot{V}(x,k(x))\leq-\gamma(V(x))
\end{equation}
where
\begin{equation}\label{eq:defVdot}
    \Dot{V}(x,k(x))=L_fV(x)+L_gV(x)k(x)
\end{equation}
then the system \cref{eq:controlaffinesystem} is stabilizable to $V(x^*)=0$, meaning $x^*=0$. In \cref{eq:defstabcontrol}, $\gamma$ is a class-$\mathcal{K}$ function defined such that $\gamma(0)=0$ and that $\gamma$ is strictly increasing. The function $V$ is known as a Lyapunov function, and it can be found that when $V\leq-\gamma(V)$ the system \cref{eq:controlaffinesystem} is stable under the control law $u=k(x)$ \cite{ames_control_2014}.

The above definitions lead to the description of a control Lyapunov function (CLF) which is a function $V$ that stabilizes a system without the need to actually construct the feedback controller $u$. All that needs to be done is to find a controller that obeys the inequality in \cref{eq:defstabcontrol}. This definition can be written as
\begin{equation}\label{eq:defofstablecontrol}
    \inf_{u\in U} [L_fV(x) +L_gV(x)]\leq -\gamma(V(x))
\end{equation}
which allows the set of all stabilizing controllers to be defined as
\begin{equation}\label{eq:setofstablecontrollers}
    K_\mathrm{clf}(x):=\{u\in U : L_fV(x)+L_gV(x)u\leq -\gamma(V(x))\}
\end{equation}
The constraint defined in \cref{eq:setofstablecontrollers} is affine in the control input $u$ meaning that it can be used for the formulation of optimization based controllers. 

\subsection{Application of CLF Theory to Traffic PDEs}
For the LWR model, the state is now a function of both space $z$ and time $t$, so \cref{eq:controlaffinesystem} becomes
\begin{equation}\label{eq:controlaffinesystem_PDE}
    \frac{\partial x}{\partial t}(z,t) = f(x(z,t)) + g(x(z,t))u
\end{equation}
where
\begin{equation}\label{eq:LWRascontorlaffinesys}
    \begin{aligned}
        f(x(z,t)) &= \mathcal{V}\frac{\partial x}{\partial z}(z,t) + \mathcal{M}x(z,t)\\
        g(x(z,t)) &= B
    \end{aligned}
\end{equation}

An admissible CLF is chosen as the quadratic deviation from a desired profile such as
\begin{equation}\label{eq:simpleCLF}
    V(t) = \frac{1}{2}\int_0^L (\Delta\rho - \Delta\rho_\mathrm{des})^2 \:dz
\end{equation}
where $\Delta\rho_\mathrm{des} \in\mathbb{R}^n$ is the desired density profile. A change of variables is introduced by defining the error system as
\begin{equation}\label{eq:errorvariable}
    e = \Delta\rho - \Delta\rho_\mathrm{des}
\end{equation}
meaning
\begin{equation}\label{eq:errorLWRsystem}
    \begin{aligned}
        \frac{\partial e}{\partial t} &= f(e) + g(e)u\\
        f(e) &= \mathcal{V}\frac{\partial e}{\partial z}(z,t) + \mathcal{M}e(z,t)\\
        g(e) &= B
    \end{aligned}
\end{equation}
The CLF defined in \cref{eq:simpleCLF} then becomes
\begin{equation}\label{eq:simpleerrorCLF}
    V(t) = \frac{1}{2}\int_0^L e^2 \:dz
\end{equation}
The dynamics of the CLF are defined as
\begin{equation}\label{eq:CLFdynamics}
    \begin{aligned}
        \Dot{V}(t) &= \int_0^L e \cdot \frac{\partial}{\partial t}e \:dz\\
        &= \int_0^L e(f(e)+g(e)u) \:dz\\
        &= \int_0^L ef(e) \:dz + \int_0^L eg(e)u \:dz\\
        &= \int_0^L e\frac{\partial e}{\partial z} \:dz + B\int_0^L eu \:dz\\
        &= \frac{e^2}{2}\bigg|_{z=0}^{z=L} + B\int_0^L eu \:dz
    \end{aligned}
\end{equation}
Using this definition of a CLF, the following quadratic programming controller is considered
\begin{equation}\label{eq:CLFQPoptprob}
    \begin{aligned}
        u(z,t) = \argmin_{(u,\delta)\in\mathbb{R}^{(n+1)\times(m+1)}} &\frac{1}{2} u^THu + p\delta^2\\
        \mathrm{s.t.} \quad & L_fV + L_gVu\leq-\gamma(V)+\delta
    \end{aligned}
\end{equation}
where $H$ is any positive definite matrix, $\delta$ is a relaxation variable that ensures \cref{eq:CLFQPoptprob} is solvable, and $p>0$ penalizes the use of $\delta$.

\section{SAFE CONTROL OF TRAFFIC MODELS}\label{sec:CBF}

\subsection{Control Barrier Function Theory for ODEs}

Safety, or safe control, of a system can be thought of as enforcing invariance of a set whereby the system does not leave a safe set of operating points, rather than driving a system to a setpoint as in stability \cite{ames_control_2014, ames_control_2017, ames_control_2019}. The safe set of interest can be defined as
\begin{equation}\label{eq:safesetconditions}
    \begin{aligned}
        \mathcal{C} &= \{x\in D\subset\mathbb{R}^m : h(x)\geq 0\}\\
        \partial\mathcal{C} &= \{x\in D\subset\mathbb{R}^m : h(x) = 0\}\\
        \mathrm{Int}(\mathcal{C}) &= \{x\in D\subset\mathbb{R}^m : h(x) > 0\}\\
    \end{aligned}
\end{equation}
where $\mathcal{C}$ is known as the safe set and $h : D\subset\mathbb{R}^m\rightarrow\mathbb{R}$ is a continuously differentiable function. This leads to the definition of safety based on forward invariance of the set $\mathcal{C}$: \emph{The set $\mathcal{C}$ is forward invariant if for every $x_0\in\mathcal{C}$, $x(t)\in\mathcal{C}$ for $x(0)=x_0$ and all $t\in\mathcal{I}(x_0)$. The system \cref{eq:controlaffinesystem} is safe with respect to the set $\mathcal{C}$ if the set $\mathcal{C}$ is forward invariant} \cite{ames_control_2019}.

The continuously differentiable function $h$ is a control barrier function (CBF) if there exists an extended class $\mathcal{K}_\infty$ function $\alpha$ such that for \cref{eq:controlaffinesystem}
\begin{equation}\label{eq:defofsafecontrol}
    \sup_{u\in U} [L_fh(x) + L_gh(x)u]\geq -\alpha(h(x))
\end{equation}
for all $x\in D$. An extended class-$\mathcal{K}_\infty$ function is a function $\alpha : \mathbb{R}\rightarrow\mathbb{R}$ that is strictly increasing with $\alpha(0)=0$ and is defined on the entire real line $\mathbb{R}=(-\infty,\infty)$.

Thus, the set of all control values that render $\mathcal{C}$ safe is
\begin{equation}\label{eq:setofsafecontrol}
    K_\mathrm{cbf} = \{u\in U : L_fh(x) + L_gh(x)u + \alpha(h(x)) \geq 0\}
\end{equation}
The main result from \cite{ames_control_2019} is that if $h$ is a control barrier function on $D$ and $\frac{\partial h}{\partial z}\neq 0$ for all $x\in\partial\mathcal{C}$, then any Lipschitz controller $u(x)\in K_\mathrm{cbf}(x)$ for \cref{eq:controlaffinesystem} renders the set $\mathcal{C}$ safe and the set $\mathcal{C}$ is asymptotically stable in $D$. An additional benefit of this design is robustness to modeling errors. Although such errors might force trajectories to  leave the set $\mathcal{C}$, the asymptotic stability of $\mathcal{C}$  under any controller in $K_\mathrm{cbf}$ ensures return to the safe set $\mathcal{C}$ \cite{ames_control_2019}.

\subsection{Application of CBF Theory to Traffic PDEs}
For PDE control, the safe set $\mathcal{C}$ and CBF becomes spatially varying as well as temporally varying as the state is both time and space dependent. The set $\mathcal{C}$ is now defined as
\begin{equation}\label{eq:safesetconditionsPDE}
    \begin{aligned}
        \mathcal{C} &= \{x\in D\subset\mathbb{R}^{n\times m} : h(x)\geq 0\}\\
        \partial\mathcal{C} &= \{x\in D\subset\mathbb{R}^{n\times m} : h(x) = 0\}\\
        \mathrm{Int}(\mathcal{C}) &= \{x\in D\subset\mathbb{R}^{n\times m} : h(x) > 0\}\\
    \end{aligned}
\end{equation}
An admissible spatially varying CBF (sCBF) is considered here
\begin{equation}\label{eq:LQRCBF}
    h(x) := \lambda(z,t) - x(z,t) \geq 0
\end{equation}
where $\lambda(z,t)$ is a desired upper bound on traffic density. The conditions in \cref{eq:setofsafecontrol} are incorporated into the optimization problem \cref{eq:CLFQPoptprob} to yield a stabilizing and safe controller as
\begin{equation}\label{eq:CLFCBFQPoptprob}
    \begin{aligned}
        u(z,t) = \argmin_{(u,\delta)\in\mathbb{R}^{(n+1)\times(m+1)}} &\frac{1}{2} u^THu + p\delta^2\\
        \mathrm{s.t.} \quad & L_fV + L_gVu\leq-\gamma(V)+\delta\\
                            & L_fh + L_ghu\leq-\alpha(h)\\
                            & u_\mathrm{min}\leq u \leq u_\mathrm{max}
    \end{aligned}
\end{equation}
where $u_\mathrm{min}$ and $u_\mathrm{max}$ are the lower and upper bounds  on the control input, respectively.

\section{SIMULATION RESULTS}\label{sec:Results}
\subsection{Case Study Definition}
\begin{figure}[ht!]
    \centering
    \begin{subfigure}{0.45\columnwidth}
            \centering
            \includegraphics[width=\textwidth]{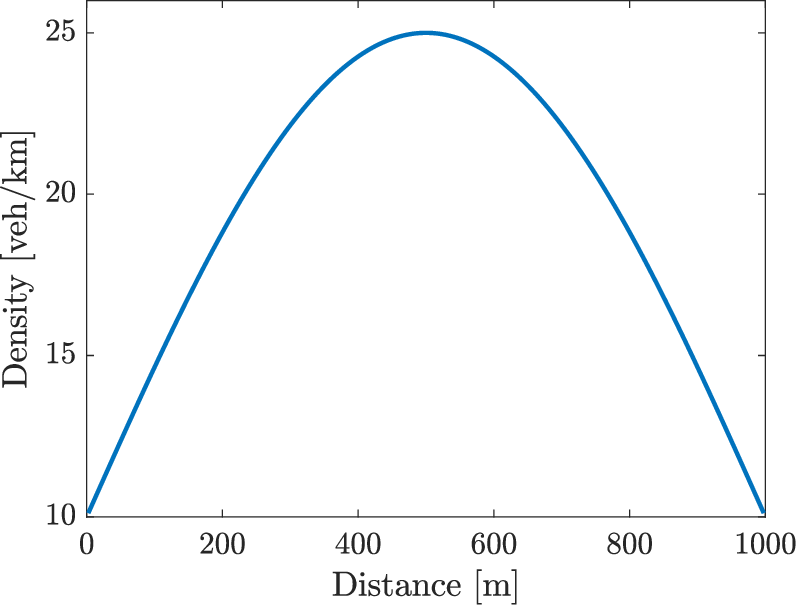}
            \caption{}
            \label{fig:IC}
    \end{subfigure}
    \begin{subfigure}{0.45\columnwidth}
            \centering
            \includegraphics[width=\textwidth]{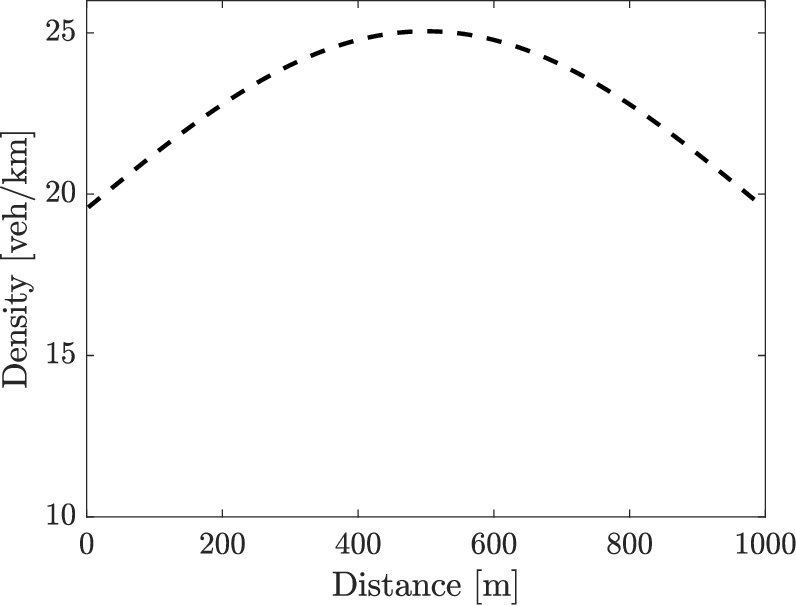}
            \caption{}
            \label{fig:lambda}
    \end{subfigure}
    \caption{Initial density profile (a) and $\lambda$ (b)}
    \label{fig:ICandCBF}
\end{figure}

A circular road with length $L = 1000$ m is chosen as the case study. In a circular road, the boundary conditions are periodic meaning that $\rho(0,t)=\rho(L,t)$. This simplifies the dynamics of the CLF to
\begin{equation}\label{eq:finalCLFdynamics}
    \begin{aligned}
        \Dot{V}(t) &= \frac{e^2}{2}\bigg|_{z=0}^{z=L} + B\int_0^L eu \:dz\\
        &= B\int_0^L eu \:dz
    \end{aligned}
\end{equation}
The parameters of the LWR model, \cref{eq:LWR,eq:fd}, are $\rho_\mathrm{max}=150$ veh/km, $\rho_\mathrm{cr}=15$ veh/km, and $V_\mathrm{max}=72$ km/h. The simulation time is $T=200$ s. The initial condition and the chosen function for $\lambda$ are shown in \cref{fig:ICandCBF}. The initial condition is completely within the safe set and includes traffic in both the free flow and congested traffic regimes.

\subsection{Results}
\begin{figure}[ht!]
    \centering
    \begin{subfigure}{0.45\columnwidth}
            \centering
            \includegraphics[width=\textwidth]{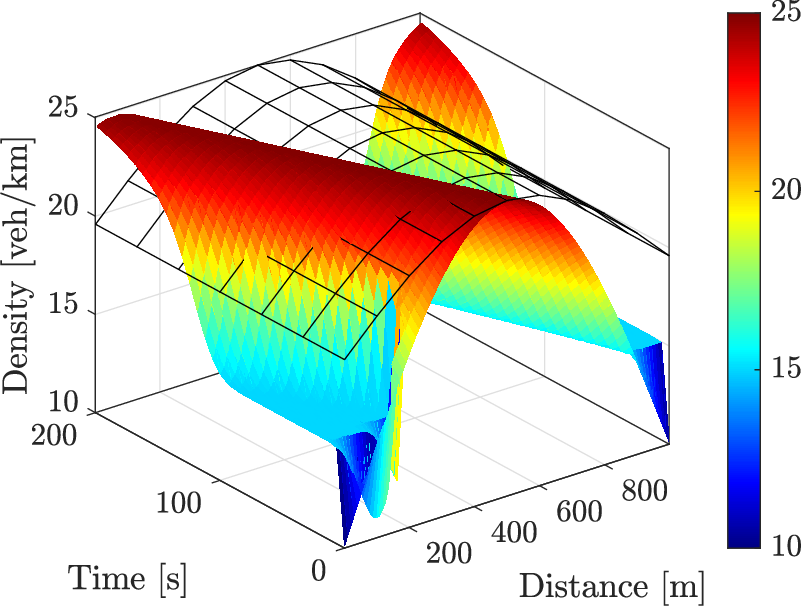}
            \caption{}
            \label{fig:UncRho}
    \end{subfigure}
    \begin{subfigure}{0.45\columnwidth}
            \centering
            \includegraphics[width=\textwidth]{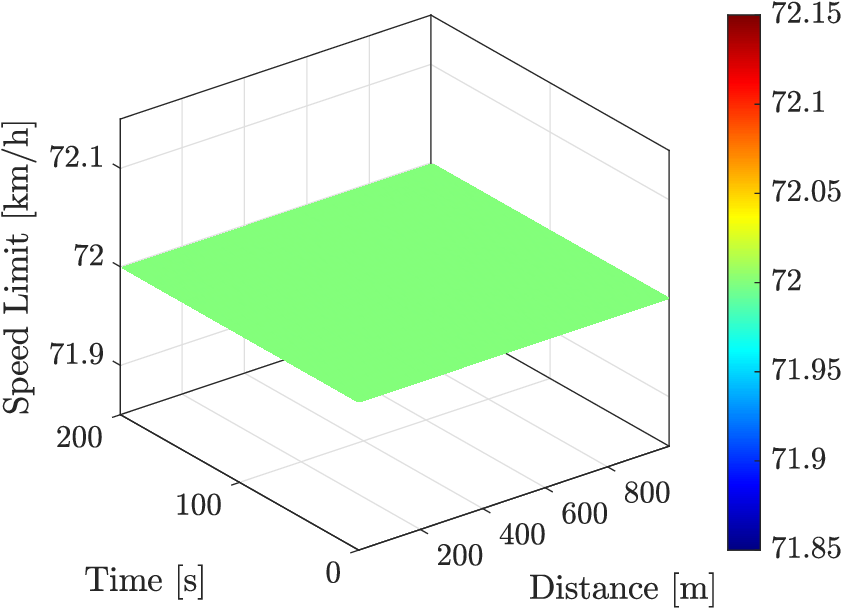}
            \caption{}
            \label{fig:UncSpdLim}
    \end{subfigure}
    \caption{Uncontrolled density (a) and speed limit (b).}
    \label{fig:UncontrolledResults}
\end{figure}

The uncontrolled circular road scenario is shown in \cref{fig:UncontrolledResults}. As can be seen in \cref{fig:UncRho}, the backwards propagating congested traffic wave violates the boundary, shown as the black mesh.

\begin{figure}[ht!]
    \centering
    \begin{subfigure}{0.45\columnwidth}
            \centering
            \includegraphics[width=\textwidth]{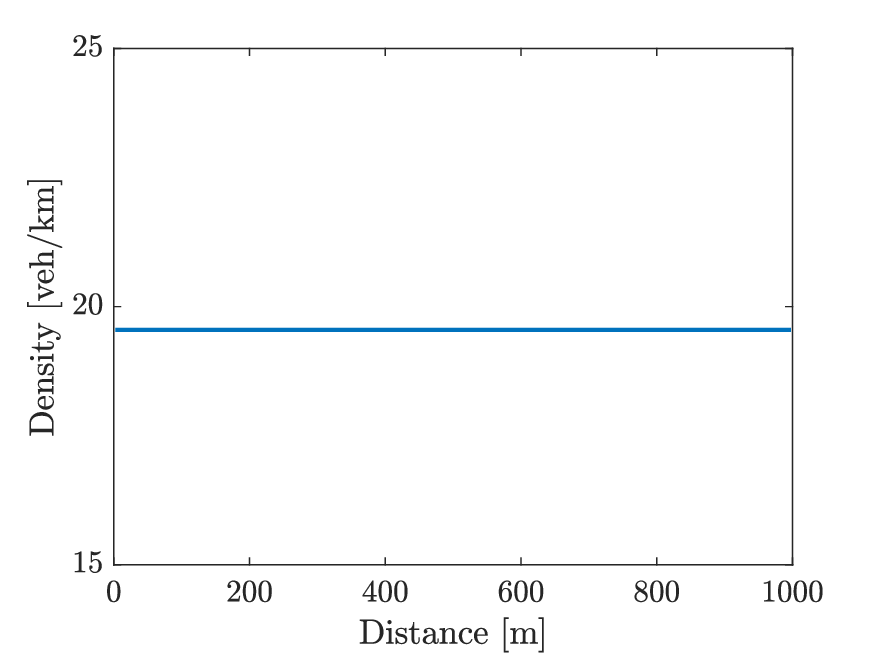}
            \caption{}
            \label{fig:rho_des1}
    \end{subfigure}
    \begin{subfigure}{0.45\columnwidth}
            \centering
            \includegraphics[width=\textwidth]{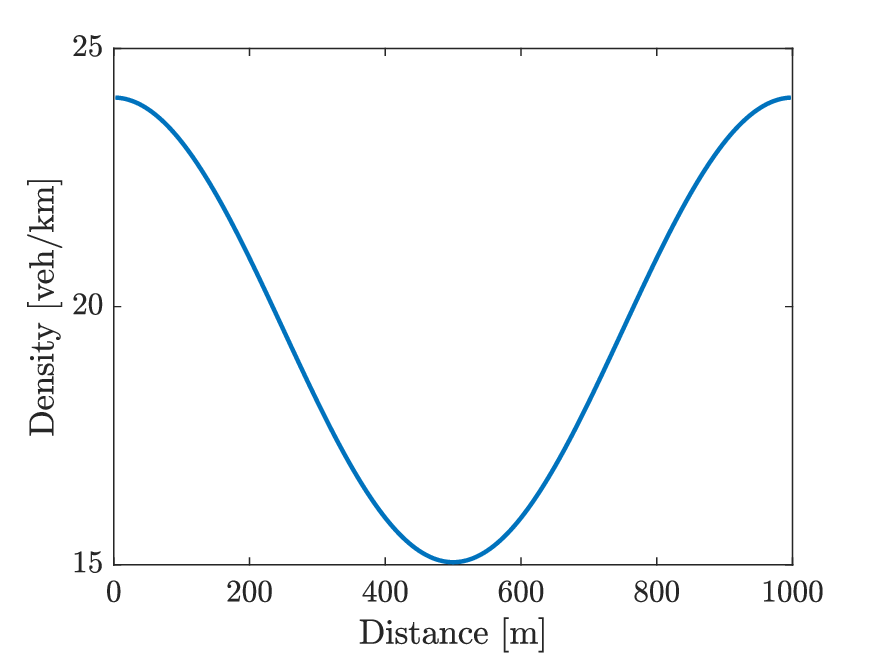}
            \caption{}
            \label{fig:rho_des2}
    \end{subfigure}
    \caption{Desired density profiles for the CLF-based controller.}
    \label{fig:rho_des}
\end{figure}

For the controlled results, three different control scenarios are examined: only stabilizing control via CLF control, only safe control via sCBF control, and combined stabilizing and safe control via CLF+sCBF control. The goal of the CLF is to drive the system to the desired density profile. Two desired profiles are used, both shown in \cref{fig:rho_des}. The profile in \cref{fig:rho_des1} falls completely within the safe region, while the profile in \cref{fig:rho_des2} drives the system outside the safe region to evaluate how robust the CLF+sCBF controller is. 

The stabilizing control solves \cref{eq:CLFQPoptprob} while the combined safe and stabilizing control solves \cref{eq:CLFCBFQPoptprob}. The safe controller solves \cref{eq:CLFCBFQPoptprob} subject only to the sCBF constraints. The class-$\mathcal{K}$ function for the CLF is chosen to be $\gamma(V) = \gamma\cdot V$ and the class-$\mathcal{K}_\infty$ in the sCBF is chosen to be $\alpha(h) = \alpha\cdot h$. The values of $\gamma$ and $\alpha$ are chosen to be 0.025 and 1, respectively. The positive definite matrix $H=I\in\mathbb{R}^{n\times m}$ and $p=1000$.

\begin{figure}[ht!]
    \centering
    \begin{subfigure}{0.45\columnwidth}
            \centering
            \includegraphics[width=\textwidth]{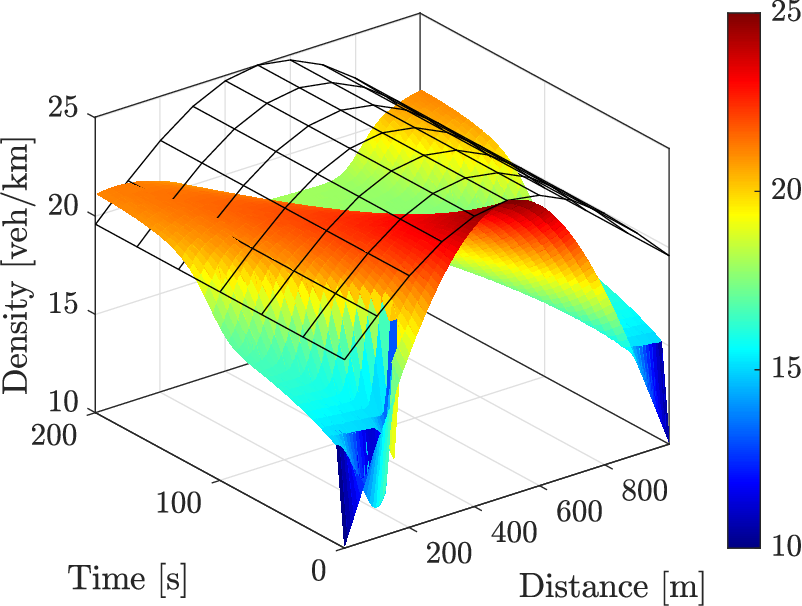}
            \caption{}
            \label{fig:CLFRho}
    \end{subfigure}
    \begin{subfigure}{0.45\columnwidth}
            \centering
            \includegraphics[width=\textwidth]{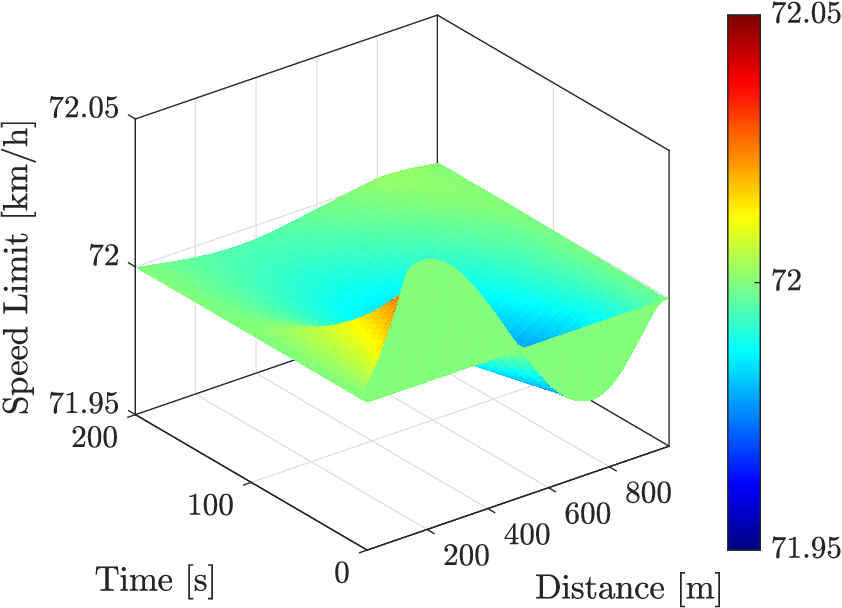}
            \caption{}
            \label{fig:CLFSpdLim}
    \end{subfigure}
    \begin{subfigure}{0.45\columnwidth}
            \centering
            \includegraphics[width=\textwidth]{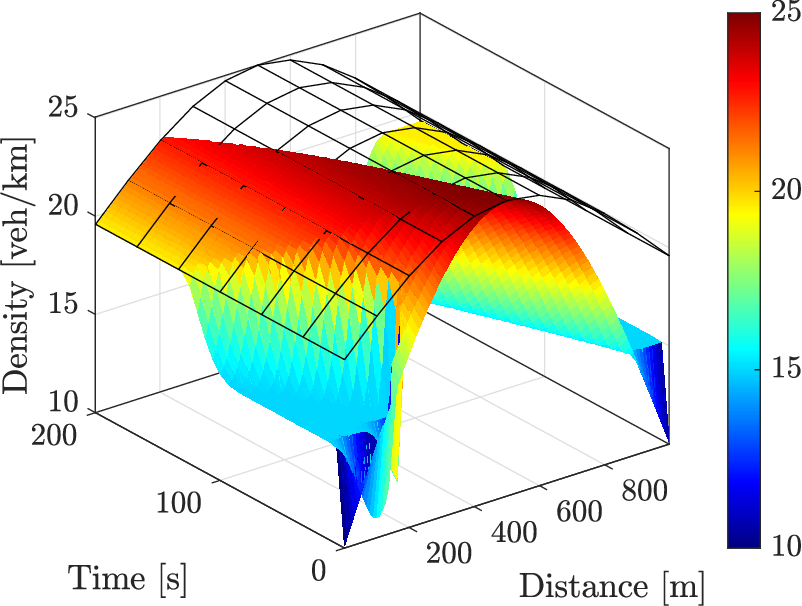}
            \caption{}
            \label{fig:CBFRho}
    \end{subfigure}
    \begin{subfigure}{0.45\columnwidth}
            \centering
            \includegraphics[width=\textwidth]{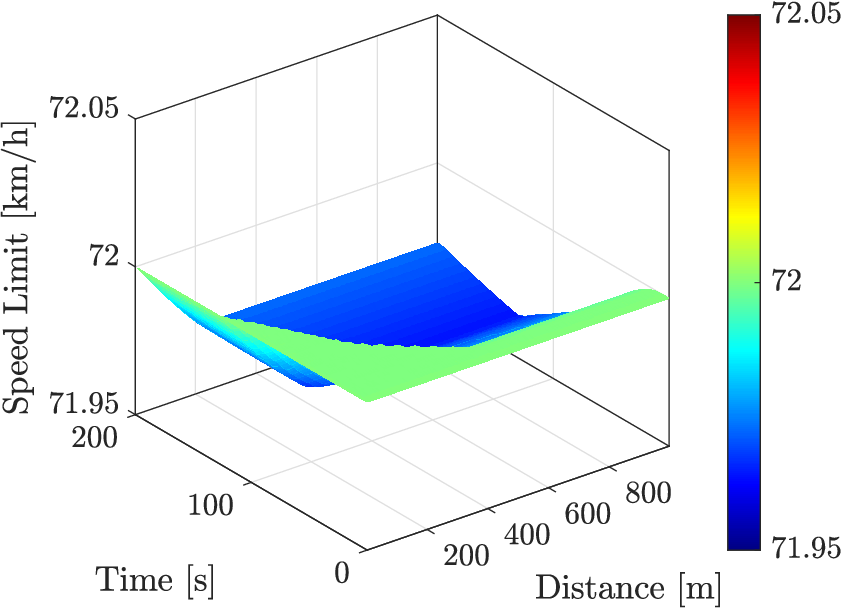}
            \caption{}
            \label{fig:CBFSpdLim}
    \end{subfigure}
    \begin{subfigure}{0.45\columnwidth}
            \centering
            \includegraphics[width=\textwidth]{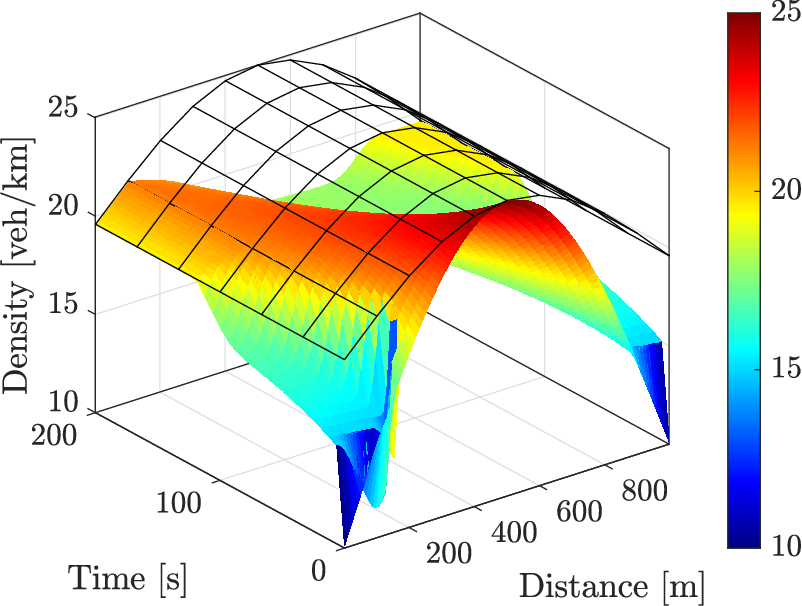}
            \caption{}
            \label{fig:CLFCBFRho}
    \end{subfigure}
    \begin{subfigure}{0.45\columnwidth}
            \centering
            \includegraphics[width=\textwidth]{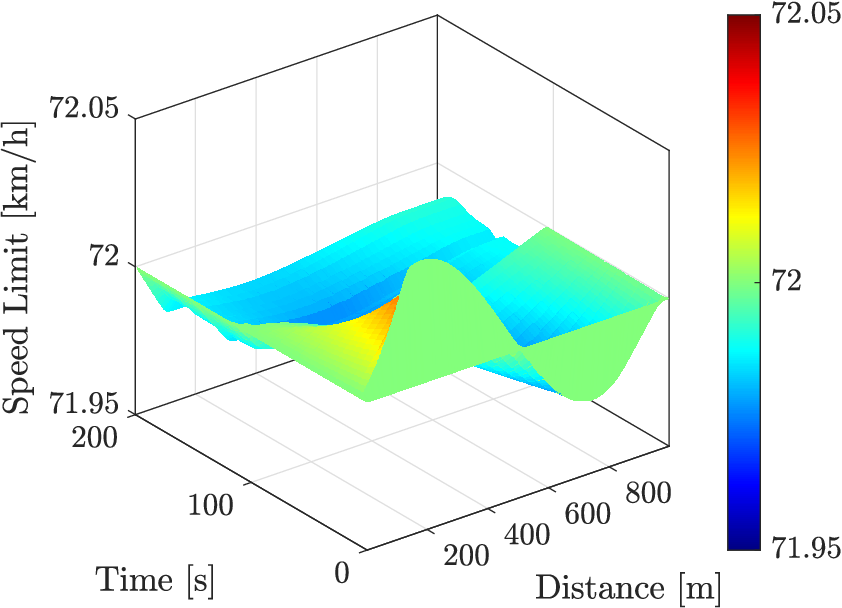}
            \caption{}
            \label{fig:CLFCBFSpdLim}
    \end{subfigure}
    \caption{Density and VSL results for CLF control (a)(b), sCBF control (c)(d), and CLF+sCBF control (e)(f).}
    \label{fig:ControlResults}
\end{figure}

The results of all three control scenarios are shown in \cref{fig:ControlResults} for the desired profile shown in \cref{fig:rho_des1}. 
When just the stabilizing controller based on the CLF is used, it start to drive the density to the desired value, but it begins to violate the boundary starting at around 100 s, as seen in \cref{fig:CLFRho}. The VSL control input is shown in \cref{fig:CLFSpdLim}. It mostly actuates in the beginning and both raises and lowers the speed limit. On the other hand, when just the sCBF is used, the speed limit is only lowered, as shown in \cref{fig:CBFSpdLim}. This results in a density profile that does not exceed $\lambda$, but also does not drive the system to the desired density. In \cref{fig:CBFRho} there is still a large deviation between maximum and minimum density.

The combined stabilizing and safe control input is shown in \cref{fig:CLFCBFSpdLim}. In the beginning, the CLF does more work as the VSL in \cref{fig:CLFCBFSpdLim} appears very similar to that in \cref{fig:CLFSpdLim}. Then, later in the simulation, the influence of the sCBF takes over and it forces a reduction in speed limit. The resulting density profile for this control input is shown in \cref{fig:CLFCBFRho}, where it can be seen that it both respects the barrier and drives the system close to the desired density. 

\begin{figure}
    \centering
    \includegraphics[width=\columnwidth]{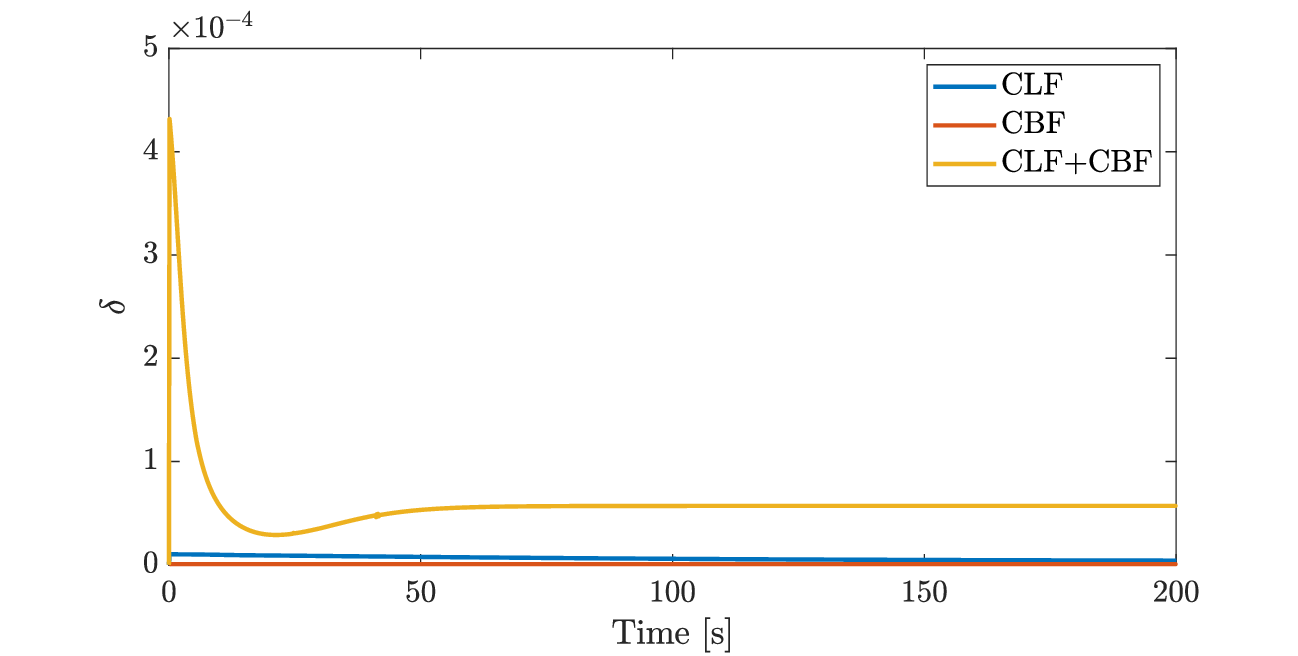}
    \caption{Relaxation variable $\delta$ for the quadratic programming problem.}
    \label{fig:delta}
\end{figure}

The value of the relaxation variable $\delta$ from the optimization problem in \cref{eq:CLFCBFQPoptprob} is reported in \cref{fig:delta} for the cases in which it is active. It can be seen that it is used more for the controller which utilizes both the CLF and the sCBF.

\begin{figure}[ht!]
    \centering
    \begin{subfigure}{0.45\columnwidth}
            \centering
            \includegraphics[width=\textwidth]{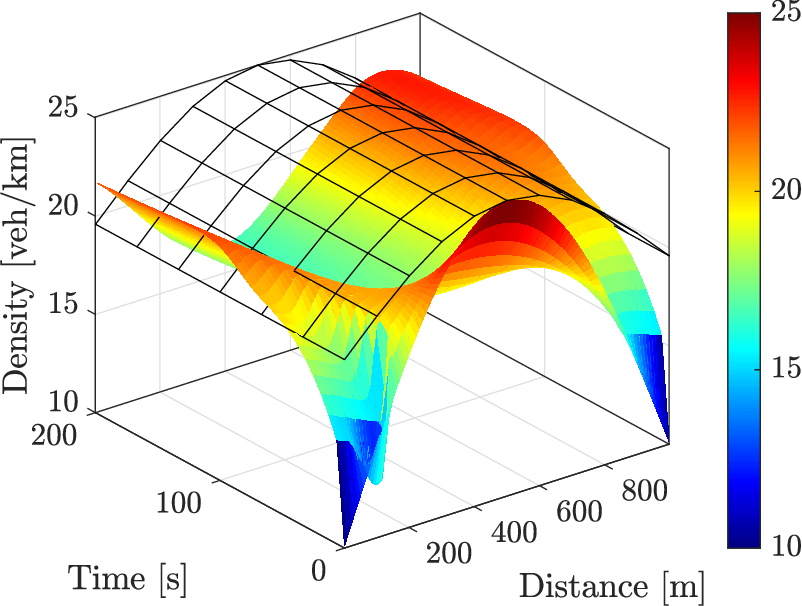}
            \caption{}
            \label{fig:CLFRho2}
    \end{subfigure}
    \begin{subfigure}{0.45\columnwidth}
            \centering
            \includegraphics[width=\textwidth]{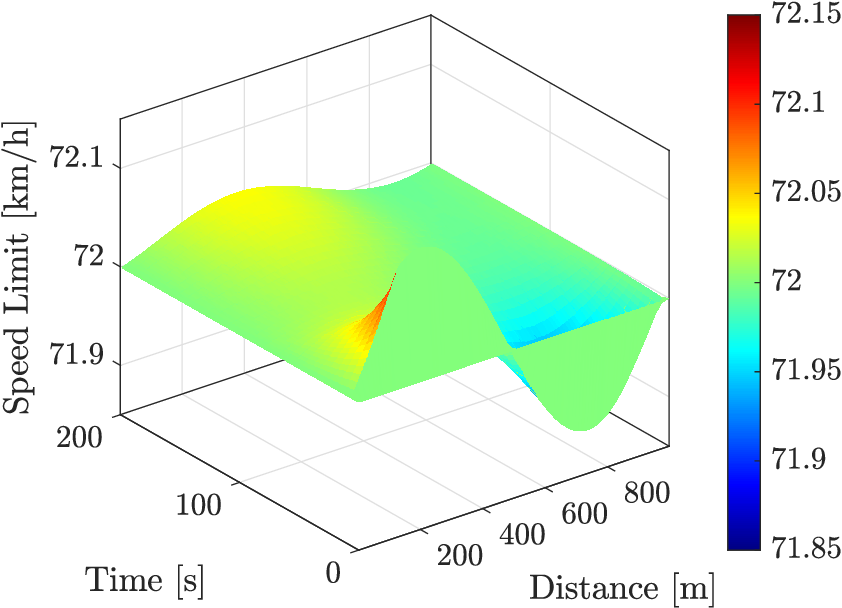}
            \caption{}
            \label{fig:CLFSpdLim2}
    \end{subfigure}
    \begin{subfigure}{0.45\columnwidth}
            \centering
            \includegraphics[width=\textwidth]{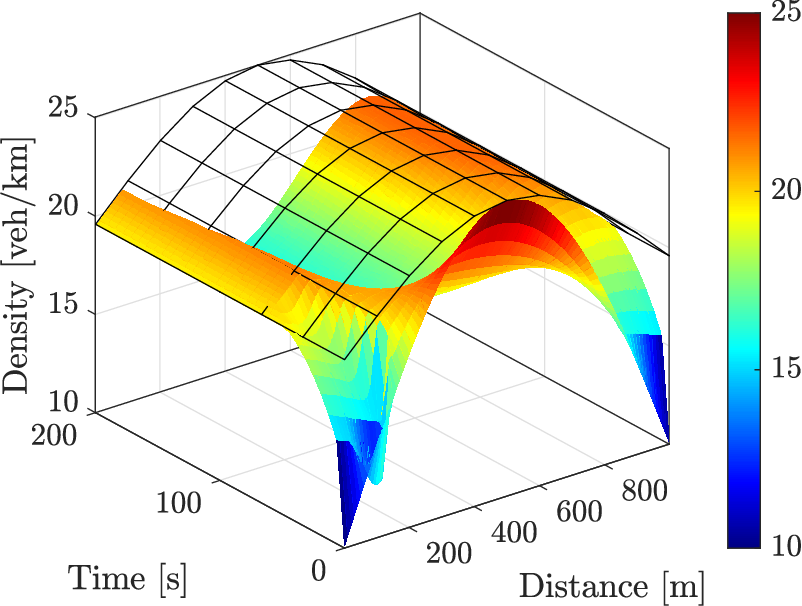}
            \caption{}
            \label{fig:CLFCBFRho2}
    \end{subfigure}
    \begin{subfigure}{0.45\columnwidth}
            \centering
            \includegraphics[width=\textwidth]{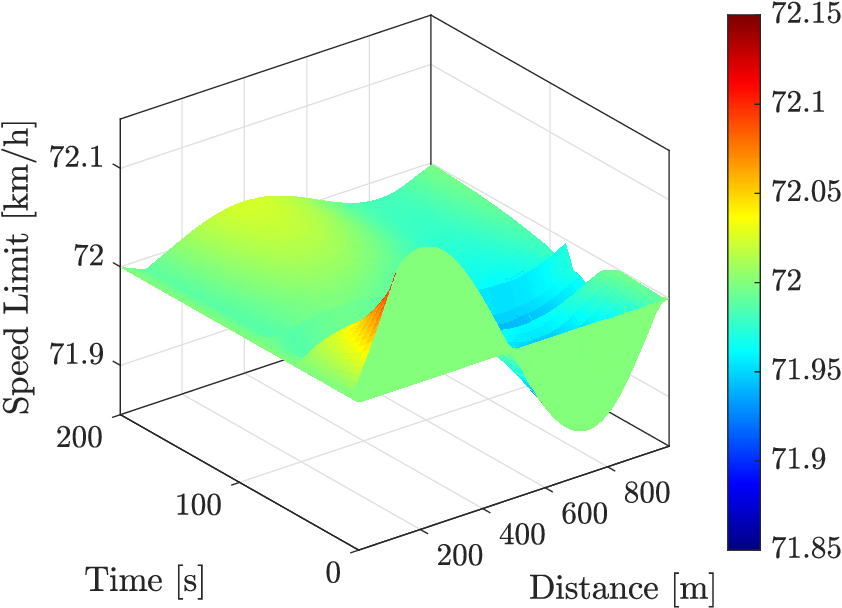}
            \caption{}
            \label{fig:CLFCBFSpdLim2}
    \end{subfigure}
    \caption{Density and VSL results for CLF control (a)(b) and CLF+CBF control (c)(d) when the desired density profile lies outside the safe set.}
    \label{fig:ControlResults2}
\end{figure}

When the desired profile is instead given by \cref{fig:rho_des2}, the resulting density and speed limits are given in \cref{fig:ControlResults2}. Since the safe controller using just the sCBF does not depend on the desired profile, there is no change so it is not shown again. As expected, the CLF-based controller drives the system close to the desired density profile and violates the barrier as seen in \cref{fig:CLFRho2} and \cref{fig:CLFSpdLim2}. When the sCBF is added in as a constraint, the controller drives the system as close as it can to the desired profile without violating the barrier as seen in \cref{fig:CLFCBFRho2}. The speed limit in this case actuates more in the region where it previously violated the barrier as the constraint is now active.

\section{CONCLUSION}
In this paper, the theory of control Lyapunov functions and control barrier functions was extended to in-domain control of traffic PDE models. First, a stabilizing controller was designed based on control Lyapunov theory. Then, control barrier function theory was extended to spatially defined functions, enabling the enforcement of safety constraints in PDE-based formulations. A unified optimization framework incorporating both CLF and sCBF constraints was then formulated. The approaches developed were applied to variable speed limit control of the LWR traffic model. Future work will focus on developing sCBFs that also depend on the control input and vehicle speed, as well as applying the constrained control method to optimal controller designs.


\bibliographystyle{IEEEtran}
\bibliography{references}

@inproceedings{xiao_control_2019,
	title = {Control {Barrier} {Functions} for {Systems} with {High} {Relative} {Degree}},
	doi = {10.1109/CDC40024.2019.9029455},
	urldate = {2024-05-13},
	booktitle = {2019 {IEEE} 58th {Conference} on {Decision} and {Control} ({CDC})},
	author = {Xiao, Wei and Belta, Calin},
	month = dec,
	year = {2019},
	pages = {474--479},
}

@article{sabouni_optimal_2024,
	title = {Optimal control of connected automated vehicles with event/self-triggered control barrier functions},
	volume = {162},
	issn = {0005-1098},
	doi = {10.1016/j.automatica.2024.111530},
	urldate = {2024-05-01},
	journal = {Automatica},
	author = {Sabouni, Ehsan and Cassandras, Christos G. and Xiao, Wei and Meskin, Nader},
	month = apr,
	year = {2024},
	pages = {111530},
}

@article{wieland_constructive_2007,
	series = {7th {IFAC} {Symposium} on {Nonlinear} {Control} {Systems}},
	title = {{C}ONSTRUCTIVE {S}AFETY {U}SING {C}ONTROL {B}ARRIER {F}UNCTIONS},
	volume = {40},
	issn = {1474-6670},
	doi = {10.3182/20070822-3-ZA-2920.00076},
	number = {12},
	urldate = {2023-11-04},
	journal = {IFAC Proceedings Volumes},
	author = {Wieland, Peter and Allgöwer, Frank},
	month = jan,
	year = {2007},
	pages = {462--467},
}

@article{karafyllis_global_2022,
	title = {Global {Stabilization} of {Compressible} {Flow} between {Two} {Moving} {Pistons}},
	volume = {60},
	issn = {0363-0129},
	doi = {10.1137/21M1413869},
	number = {2},
	urldate = {2023-10-25},
	journal = {SIAM Journal on Control and Optimization},
	author = {Karafyllis, Iasson and Krstic, Miroslav},
	month = apr,
	year = {2022},
	pages = {1117--1142},
}

@inproceedings{koga_event-triggered_2023,
	title = {Event-{Triggered} {Safe} {Stabilizing} {Boundary} {Control} for the {Stefan} {PDE} {System} with {Actuator} {Dynamics}},
	doi = {10.23919/ACC55779.2023.10156551},
	urldate = {2023-09-26},
	booktitle = {2023 {American} {Control} {Conference} ({ACC})},
	author = {Koga, Shumon and Demir, Cenk and Krstic, Miroslav},
	month = may,
	year = {2023},
	pages = {1794--1799},
}

@inproceedings{ames_control_2014,
	title = {Control barrier function based quadratic programs with application to adaptive cruise control},
	doi = {10.1109/CDC.2014.7040372},
	urldate = {2023-09-26},
	booktitle = {53rd {IEEE} {Conference} on {Decision} and {Control}},
	author = {Ames, Aaron D. and Grizzle, Jessy W. and Tabuada, Paulo},
	month = dec,
	year = {2014},
	pages = {6271--6278},
}

@article{ames_control_2017,
	title = {Control {Barrier} {Function} {Based} {Quadratic} {Programs} for {Safety} {Critical} {Systems}},
	volume = {62},
	issn = {1558-2523},
	doi = {10.1109/TAC.2016.2638961},
	number = {8},
	urldate = {2023-09-26},
	journal = {IEEE Transactions on Automatic Control},
	author = {Ames, Aaron D. and Xu, Xiangru and Grizzle, Jessy W. and Tabuada, Paulo},
	month = aug,
	year = {2017},
	pages = {3861--3876},
}

@inproceedings{ames_control_2019,
	title = {Control {Barrier} {Functions}: {Theory} and {Applications}},
	shorttitle = {Control {Barrier} {Functions}},
	doi = {10.23919/ECC.2019.8796030},
	urldate = {2023-09-26},
	booktitle = {2019 18th {European} {Control} {Conference} ({ECC})},
	author = {Ames, Aaron D. and Coogan, Samuel and Egerstedt, Magnus and Notomista, Gennaro and Sreenath, Koushil and Tabuada, Paulo},
	month = jun,
	year = {2019},
	pages = {3420--3431},
}

@article{richards_shock_1956,
	title = {Shock {Waves} on the {Highway}},
	volume = {4},
	issn = {0030-364X},
	doi = {10.1287/opre.4.1.42},
	number = {1},
	urldate = {2023-05-01},
	journal = {Operations Research},
	author = {Richards, Paul I.},
	month = feb,
	year = {1956},
	pages = {42--51},
}

@article{lighthill_kinematic_1955,
	title = {On kinematic waves {II}. {A} theory of traffic flow on long crowded roads},
	volume = {229},
	issn = {0080-4630, 2053-9169},
	doi = {10.1098/rspa.1955.0089},
	
	number = {1178},
	urldate = {2023-05-01},
	journal = {Proceedings of the Royal Society of London. Series A. Mathematical and Physical Sciences},
	author = {Lighthill, M.J and Whitham, G.B.},
	month = may,
	year = {1955},
	pages = {317--345},
}

@article{koga_safe_2023,
	title = {Safe {PDE} {Backstepping} {QP} {Control} {With} {High} {Relative} {Degree} {CBFs}: {Stefan} {Model} {With} {Actuator} {Dynamics}},
	issn = {1558-2523},
	shorttitle = {Safe {PDE} {Backstepping} {QP} {Control} {With} {High} {Relative} {Degree} {CBFs}},
	doi = {10.1109/TAC.2023.3250514},
	journal = {IEEE Transactions on Automatic Control},
	author = {Koga, Shumon and Krstic, Miroslav},
	year = {2023},
	pages = {1--14},
}

@article{carlson_local_2011,
	title = {Local {Feedback}-{Based} {Mainstream} {Traffic} {Flow} {Control} on {Motorways} {Using} {Variable} {Speed} {Limits}},
	volume = {12},
	issn = {1558-0016},
	doi = {10.1109/TITS.2011.2156792},
	number = {4},
	journal = {IEEE Transactions on Intelligent Transportation Systems},
	author = {Carlson, Rodrigo Castelan and Papamichail, Ioannis and Papageorgiou, Markos},
	month = dec,
	year = {2011},
	pages = {1261--1276},
}

@inproceedings{pasquale_comparative_2016,
	title = {A comparative analysis of solution algorithms for nonlinear freeway traffic control problems},
	doi = {10.1109/ITSC.2016.7795798},
	booktitle = {2016 {IEEE} 19th {International} {Conference} on {Intelligent} {Transportation} {Systems} ({ITSC})},
	author = {Pasquale, C. and Anghinolfi, D. and Sacone, S. and Siri, S. and Papageorgiou, M.},
	month = nov,
	year = {2016},
	pages = {1773--1778},
}

@article{siri_freeway_2021,
	title = {Freeway traffic control: {A} survey},
	volume = {130},
	issn = {0005-1098},
	shorttitle = {Freeway traffic control},
	doi = {10.1016/j.automatica.2021.109655},
	
	urldate = {2022-09-21},
	journal = {Automatica},
	author = {Siri, Silvia and Pasquale, Cecilia and Sacone, Simona and Ferrara, Antonella},
	month = aug,
	year = {2021},
	pages = {109655},
}

@article{muralidharan_computationally_2015,
	series = {Special {Issue}: {Advanced} {Road} {Traffic} {Control}},
	title = {Computationally efficient model predictive control of freeway networks},
	volume = {58},
	issn = {0968-090X},
	doi = {10.1016/j.trc.2015.03.029},
	
	urldate = {2022-09-21},
	journal = {Transportation Research Part C: Emerging Technologies},
	author = {Muralidharan, Ajith and Horowitz, Roberto},
	month = sep,
	year = {2015},
	pages = {532--553},
}

@article{karafyllis_feedback_2019-1,
	title = {Feedback control of scalar conservation laws with application to density control in freeways by means of variable speed limits},
	volume = {105},
	issn = {0005-1098},
	doi = {10.1016/j.automatica.2019.03.021},
	
	urldate = {2022-09-21},
	journal = {Automatica},
	author = {Karafyllis, Iasson and Papageorgiou, Markos},
	month = jul,
	year = {2019},
	pages = {228--236},
}

@article{yu_traffic_2019,
	title = {Traffic congestion control for {Aw}–{Rascle}–{Zhang} model},
	volume = {100},
	issn = {0005-1098},
	doi = {10.1016/j.automatica.2018.10.040},
	
	urldate = {2022-06-25},
	journal = {Automatica},
	author = {Yu, Huan and Krstic, Miroslav},
	month = feb,
	year = {2019},
	pages = {38--51},
}

@article{yu_reinforcement_2022,
	title = {Reinforcement {Learning} {Versus} {PDE} {Backstepping} and {PI} {Control} for {Congested} {Freeway} {Traffic}},
	volume = {30},
	issn = {1558-0865},
	doi = {10.1109/TCST.2021.3116796},
	number = {4},
	journal = {IEEE Transactions on Control Systems Technology},
	author = {Yu, Huan and Park, Saehong and Bayen, Alexandre and Moura, Scott and Krstic, Miroslav},
	month = jul,
	year = {2022},
	note = {Conference Name: IEEE Transactions on Control Systems Technology},
	pages = {1595--1611},
}

@inproceedings{block_lq_2024,
	title = {{LQ} {Control} of {Traffic} {Flow} {Models} via {Variable} {Speed} {Limits}},
	copyright = {Creative Commons Attribution-NonCommercial-NoDerivatives 4.0 International License},
	doi = {10.23919/ACC60939.2024.10644572},
	urldate = {2024-11-06},
	booktitle = {2024 {American} {Control} {Conference} ({ACC})},
	author = {Block, Brian and Stockar, Stephanie},
	month = jul,
	year = {2024},
	pages = {4262--4267},
}

@article{ahmadi_safety_2017,
	title = {Safety verification for distributed parameter systems using barrier functionals},
	volume = {108},
	issn = {0167-6911},
	doi = {10.1016/j.sysconle.2017.08.002},
	urldate = {2025-06-23},
	journal = {Systems \& Control Letters},
	author = {Ahmadi, Mohamadreza and Valmorbida, Giorgio and Papachristodoulou, Antonis},
	month = oct,
	year = {2017},
	pages = {33--39},
}

@inproceedings{park_discretization-robust_2023,
	title = {Discretization-{Robust} {Safety} {Barrier} of {Partial} {Differential} {Equation}},
	doi = {10.1109/ICCMA59762.2023.10374980},
	urldate = {2025-06-23},
	booktitle = {2023 11th {International} {Conference} on {Control}, {Mechatronics} and {Automation} ({ICCMA})},
	author = {Park, Younghwa and Sloth, Christoffer},
	month = nov,
	year = {2023},
	pages = {49--54},
}

@inproceedings{ahmadi_barrier_2015,
	title = {Barrier functionals for output functional estimation of {PDEs}},
	doi = {10.1109/ACC.2015.7171125},
	urldate = {2025-06-23},
	booktitle = {2015 {American} {Control} {Conference} ({ACC})},
	author = {Ahmadi, Mohamadreza and Valmorbida, Giorgio and Papachristodoulou, Antonis},
	month = jul,
	year = {2015},
	pages = {2594--2599},
}

@article{block_lq-informed_2025,
	series = {11th {IFAC} {Symposium} on {Advances} in {Automotive} {Control} {AAC} 2025},
	title = {{LQ}-{Informed} {Rule}-{Based} {Variable} {Speed} {Limit} {Control}},
	volume = {59},
	copyright = {All rights reserved},
	issn = {2405-8963},
	doi = {10.1016/j.ifacol.2025.07.090},
	number = {5},
	urldate = {2025-08-11},
	journal = {IFAC-PapersOnLine},
	author = {Block, Brian and Stockar, Stephanie},
	month = jan,
	year = {2025},
	pages = {109--114},
}

@article{daganzo_cell_1994,
  title={The cell transmission model: A dynamic representation of highway traffic consistent with the hydrodynamic theory},
  author={Daganzo, Carlos F},
  journal={Transportation Research Part B: Methodological},
  volume={28},
  number={4},
  pages={269--287},
  year={1994},
  publisher={Elsevier}
}

\end{document}